\documentclass[twocolumn,amsmath,amssymb,aps]{revtex4}
\usepackage{amstext}
\usepackage{graphicx}
\usepackage{esint}

\usepackage{amstext}

\usepackage{dcolumn}
\usepackage{bm}
\usepackage[mathlines]{lineno}

\makeatother

\begin{document}
\title{Quantum random number generation by coherent detection of phase noise}
\author{Juan-Rafael \'{A}lvarez$^{1,2}$}
\email{juan.alvarezvelasquez@physics.ox.ac.uk}

\author{Samael Sarmiento$^{3}$}
\author{Jos\'{e} A. L\'{a}zaro$^{3}$}
\author{Joan M. Gen\'{e}$^{3}$}
\author{Juan P. Torres$^{1,3}$}
\affiliation{$^{1}$ICFO - Institut de Cienci\`{e}s Fot\`{o}niques, The Barcelona
Institute of Science and Technology, 08860 Castelldefels (Barcelona),
Spain~~\\
 $^{2}$University of Oxford, Clarendon Laboratory, Parks Road, Oxford
OX1 3PU, United Kingdom~~\\
 $^{3}$ Department of Signal Theory and Communications, Universitat
Polit\`{e}cnica de Catalunya, 08034 Barcelona, Spain~~ }
\date{\today}
\begin{abstract}
We demonstrate a quantum random number generator based on the random
nature of the phase difference between two independent laser sources.
The speed of random bit generation is determined by the photodetector
bandwidth and the linewidth of the lasers used. The system implemented
is robust and generates a probability distribution of quantum origin
which is intrinsically uniform and thus in principle needs no randomness
extraction. The phase is measured with telecom equipment routinely
used for high capacity coherent optical communications, which allows
to keep track of the phase drift of the lasers and is readily available
in the telecommunication industry. 
\end{abstract}
\maketitle

\section{\label{sec:Introduction}Introduction}

Random numbers are routinely needed in many branches of science and
technology. They are a key element in the development of secure communications
channels, since random keys can provide unbreakable encryption systems
\cite{Shannon1949}. They are used in banking, which uses the RSA
algorithm that relies on the generation of random numbers \cite{RSA}.
They are important in gambling, where excellent random number generators
are needed to guarantee the fairness of used machines.
Indeed, governments are implementing technical standards on the usage
of random number generators \cite{Turan2018}. In scientific applications,
Random Number Generators are behind powerful simulation methods such
as Monte Carlo \cite{Gentle}. Random numbers are generated in many
different ways. There are algorithms that generate streams of numbers
(Pseudo-Random Number generators, PRNGs) that, in spite of not being
truly random, can faithfully simulate true random sequences. There
also exist methods which are based on deterministic processes, but
for which our ignorance of the many variables involved make the possible
outcomes random. This is the case, for instance, of Random Number
Generators (RNGs) based on the behavior of chaotic systems \cite{Bonilla2016},
certain geological events such as earthquakes, astronomical events,
motion of computer mice, and interactions in social media \cite{Leung,Chen,ZhouMiceMotion}.

Randomness sources based on the principles of quantum mechanics can
provide true randomness which stems from fundamental physical phenomena
\cite{Herrero}. In principle, the generated random keys are intrinsically
random, as opposed to other random number generators based on different
physical principles which might contain biases due to phenomena which
are deterministic, yet unknown to an experimenter. Due to this, there
is a lot of interest in the generation of random numbers based on
the intrinsic and fundamental random character of quantum phenomena.

One particular phenomena that can be considered for Quantum Random
Number Generation (QRNG) is the temporal drift of the phase of any
laser source. Spontaneous emission of radiation, a quantum phenomena
whose presence is inevitable in any lasing process, is responsible
for such a phase drift. Phase noise based QRNGs, in combination with
the use of pulsed lasers, have achieved speeds of the order of 43
Gbit/s \cite{Abellan}. Lately, these phase noise-based QRNGs have
been a key element to close certain loopholes existing in Bell's inequalities
tests \cite{Giustina}.

In 2010, Qi \emph{et al.} \cite{Qi} demonstrated a phase noise based
random number generator by passing the signal of a laser by an unbalanced
Mach-Zehnder interferometer. The idea behind this experiment is interfering
the laser signal at two different times so that the phase difference
between these two signals is known to evolve randomly. The experimental
setup for this situation is shown in Figure \ref{fig:EquivalenceFigure}
(a). If $\tau_{\text{coh}}$ is the coherence time of the laser, $T_{\text{det}}$
is the response time of the photodetectors, and $T_{\text{sample}}$
is the time an Analog to Digital Converter (ADC) takes between successive
measurements, three conditions should be fulfilled for Random Number
Generation: 
\begin{enumerate}
\item $T_{\text{delay}}\gg\tau_{\text{coh}}$ to guarantee that the phases
in both arms of the interferometer are uncorrelated. 
\item $T_{\text{sample}}>\tau_{\text{ coh }}$ to take samples after the
phase difference has effectively changed. 
\item $T_{\text{det}}<\tau_{\text{coh}}$ to detect a fixed phase and make
sure we are not integrating over a phase change. 
\end{enumerate}
These conditions pose limitations on experimental realizations of
this QRNG, as time delays of the order of $\tau_{c}$ imply long path
delays. Such delays are usually implemented using optical fiber cables.
As an example, a coherence time of 0.01 milliseconds (a linewidth
of 100 kHz), would require 3 additional kilometers of fiber in one
of the interferometer arms.

The presence of long delays pose challenges due to the losses that
fibres present, especially in non-telecom wavelengths, as well as
their difficulty for possible integration in chips. One way to circumvent
this drawback is using a slightly unbalanced interferometer that would
generate a random phase with a non-uniform Gaussian-like probability
distribution\cite{Qi}. However, this calls for randomness extraction
and the active stabilization of the interferometer that increases
the technical difficulty of the implementation of the system.

Here we put forward and demonstrate a new scheme where the quantum
origin of the phase noise can be exploited for the generation of random
number sequences without the need of using a highly unbalanced interferometer
with long delays or phase stabilization in a slightly unbalanced interferometer.
We use a second laser and measure the interference of the two laser
sources with a coherent detector (homodyne detection) that measure
the phase variation directly. This method, which readily exposes the
random behavior of the phase, requires a minimal amount of post-processing
from the measured signal, since the probability distribution of the
phase is uniform. Homodyne detection of optical signals is routinely
done in optical communications laboratories with the help of optical
hybrids. In optical communications, one is interested in extracting
the real and imaginary part of a signal, which carries the information,
with the help of a local oscillator that acts as reference signal.
For random number generation, we are interested in extracting the
phase difference between two signals.

In particular, we perform balanced coherent detection of two narrowband
laser signals in the telecom band generated with external cavity lasers.
By using a fast sampling of the order of 100 MSamples/s, we are able
to characterize the random walk performed by their phase difference,
and determine their coherence times, i.e., the time over which the
phases of the two lasers can be considered constant. However, by using slow
sampling, of the order of 100 kSamples/s, we are able to generate random numbers which do not require
any sort of randomness extraction and thus are ready to be used after
being measured. We will show that we do not require randomness extraction
to pass almost all of the standard randomness tests designed for RNGs.
This is a feature that can be used for future, faster QRNGs, as we
are able to provide true randomness with keys that are fast enough
to feed randomness extractors with nonuniform distributions of quantum
origin.

\section{Methods}

\begin{figure*}
\centering{}\centering{}\includegraphics[width=2\columnwidth]{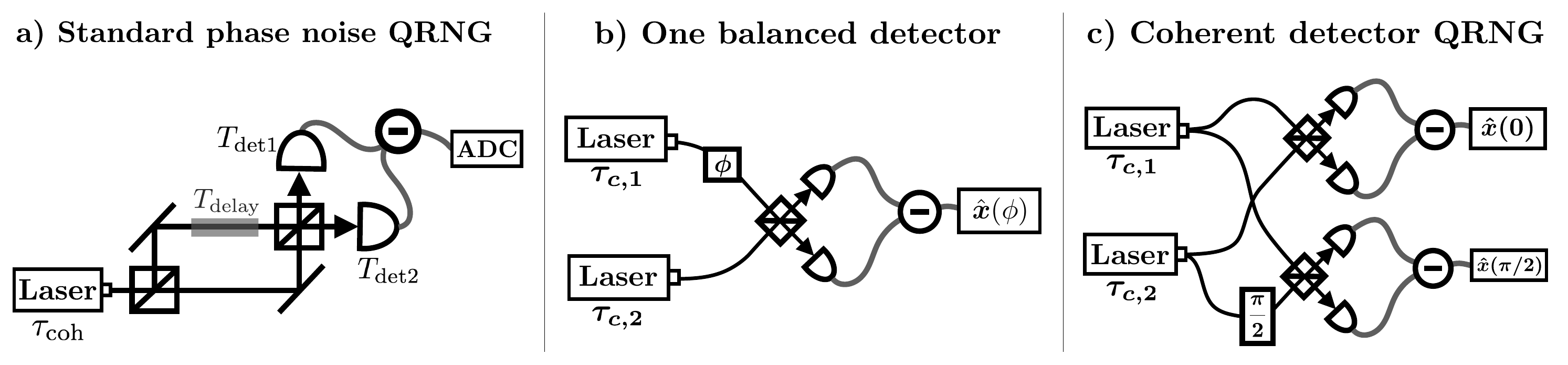}\caption{\label{fig:EquivalenceFigure} (a) Depiction of an experimental setup
for a phase noise interference Quantum Random Number generator, in
which a laser interferes with itself at points where the phase
has changed. (b) Schematic of a balanced detector, in which a phase
$\phi$ is introduced in one of the lasers. A balanced detector with
a phase delay $\phi$ can measure the quadrature $\hat{x}\left(\phi\right)$
of the interfering field. (c) Schematic of the setup
used for a balanced coherent detector QRNG, which measures the quadratures
of the interference between two lasers and thus obtains the phase
noise of such interference.}
\end{figure*}

The source of randomness comes from the random phase difference between
the signals coming from two different lasers. To measure such phases
we make both signals to interfere with the help of balanced coherent
detectors as shown in Figure \ref{fig:EquivalenceFigure} (b) and
(c). These are optical arrangements which involve the use of phase
delays and beam splitters. The signal of interest comes from the subtraction
of the photocurrents measured by optical detectors at the two output
ports of the beam splitter. The use of two balanced detectors with
different phase delays makes it possible to recover the complex information
of the interfering waves, in a configuration that is called a coherent
detector. QRNGs using coherent detectors and the interference of the
vacuum field with coherent fields have been described theoretically
\cite{Zhou} and demonstrated experimentally \cite{Avesani}.

Using a balanced detector with a phase delay $\phi$, it is possible
to measure the mean value of the quadrature $\hat{x}\left(\phi\right)$
of the interference between two lasers with frequencies $\omega_{1/2}$
and intensities $I_{1/2}$. This gives rise to a Skellam probability
distribution for the measurement of the quadrature $\hat{x}\left(\phi\right)$
\cite{Zhou}, which can be approximated as a normal probability distribution
with mean $\sqrt{I_{1}I_{2}}\cos\left(\Delta_{\omega}t+\phi\right)$
and variance $2\left(I_{1}+I_{2}\right)$, where $\Delta_{\omega}=\omega_{1}-\omega_{2}$.

As the phase changes in time due to spontaneous emission events occurring
in the laser cavities, the signal measured by a balanced detector
with a phase delay $\phi$ can be written as 
\begin{equation}
\left\langle x\left(\phi\right)\right\rangle \propto\sqrt{I_{1}I_{2}}\cos\left(\Delta_{\omega}t+\xi\left(t\right)+\phi\right),
\end{equation}
where $\xi\left(t\right)$ is a normally distributed phase with zero
mean and a variance that becomes broader as time progresses \cite{Sun2017}:
\begin{equation}
\text{var}\left(\xi\left(t\right)\right)=2t\left(\frac{1}{\tau_{c,1}}+\frac{1}{\tau_{c,2}}\right)=\frac{t}{\overline{\tau_{c}}},
\end{equation}
where $\tau_{c,1/2}$ are the coherence times of the input lasers,
defined as the inverse of their linewidths. Since $\xi\left(t\right)$
can only take values on the range $\left]-\pi,\pi\right]$, phases
are wrapped around this range and are distributed according to a von
Mises distribution\cite{Mardia}:

\begin{equation}
P\left(\xi\left(t\right)=\theta\right)=\frac{e^{\overline{\tau_{c}}\cos(\theta)/t}}{2\pi I_{0}(t/\overline{\tau_{c}})}.\label{eq:vonMises}
\end{equation}

where $I_{0}\left(x\right)$ is the zeroth-order modified Bessel function
of the first kind \cite{AbramowitzStegun}.

If the values of the phase are digitalized, the probability distribution
of the phase can be made arbitrarily similar to a uniform distribution
by measuring at times which are much longer than the coherence time.
In fact, for a digitalization of $8$ bits in $\theta$ and $t>2\overline{\tau_{c}}$,
the maximum probability difference between the von Mises distribution
and a uniform distribution is smaller than $10^{-5}$, as shown in the Supplementary Information, section A.

\section{Results}

We performed experiments with two external cavity lasers (HP 8168A
and Agilent 8164A), with wavelengths of around $1550\text{nm}$,
i.e., central frequencies $\nu_{0}\approx193.4\text{THz}$. Typical
central frequency differences measured between the two lasers are
of the order of $1\text{GHz},$ resulting in wavelength differences
of the order of $8\text{pm}$. The linewidths
of the lasers are $\Delta\nu\approx100\text{kHz}$, thus providing
coherence times of $\tau_{c}\approx20\text{\ensuremath{\mu}s}$. The
laser powers are of $P=0.1\text{mW}$ at the photodetector end, corresponding
to mean photon flux numbers of $\Phi=\left(P/h\nu_{0}\right)\approx7.8\times10^{14}\text{ photons/s}$.

We use balanced detectors (Thorlabs PDB480C-AC) with 1.6 GHz Bandwidth
($T_{\text{det}}=625\text{ps}$). With this response time, detectors
are able to measure $\approx4.9\times10^{5}\text{photons}$ per sample.
The readout is then digitalized with an oscilloscope (Tektronix MSO
70804C) at sample rates ranging from 156.25kSamples/s to 25GSamples/s.

\begin{figure}
\centering{}\includegraphics[width=0.95\columnwidth]{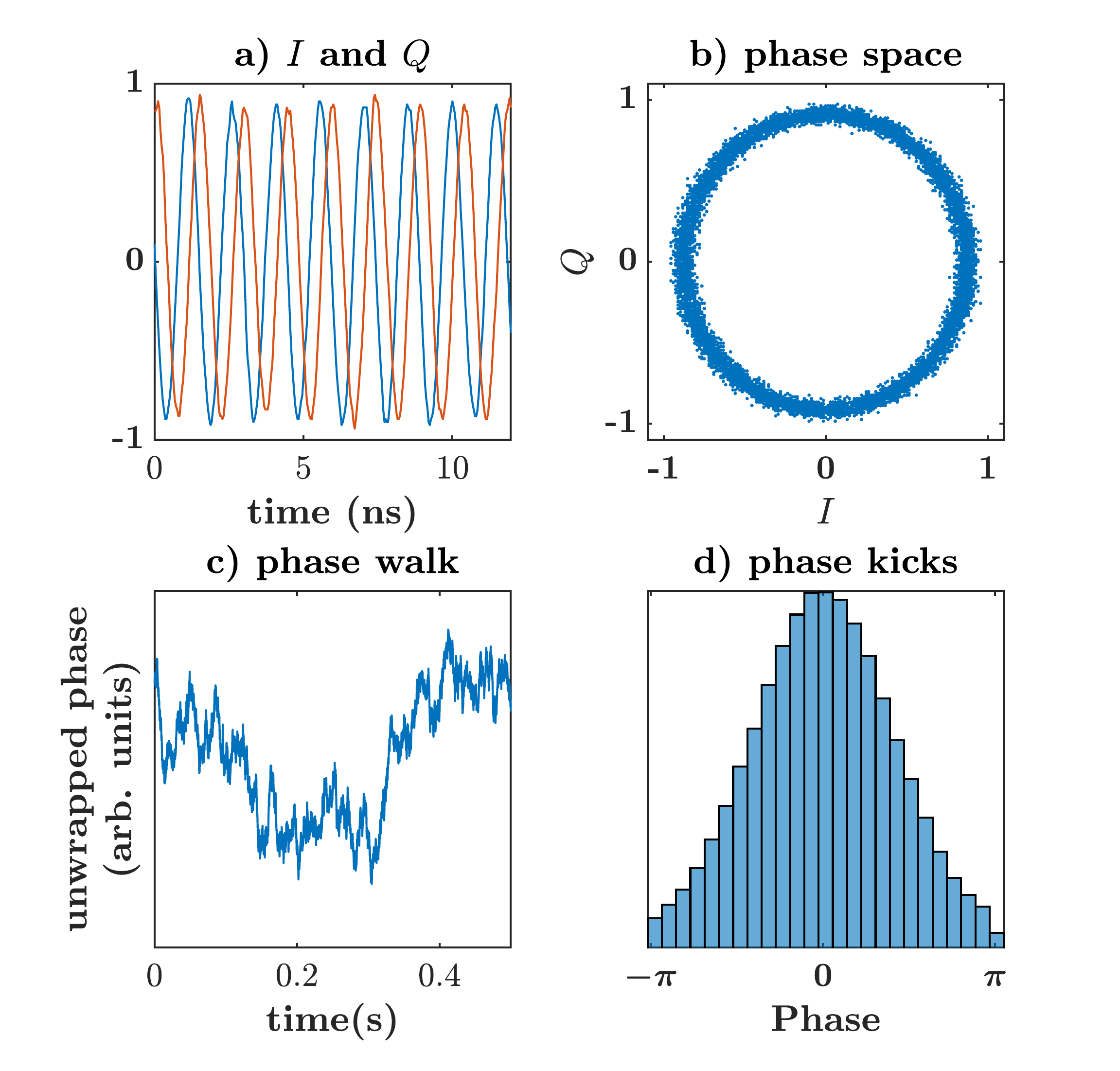}\caption{\label{fig:SpaceWalksKicks} Depiction of the processing methods to
measure the phase noise difference of the lasers. Figure \textbf{(a)}
is measured at sample rates of 25 GSamples/s, whereas figures (b-d)
are measured at sample rates of 625 kSamples/s. \textbf{(a)} Presents
an experimental measurement of the components $I$ (blue) and $Q$
(orange) as they are read out. It is noticeable that these two measurements
are offset by $\pi/2$, and indeed they draw a circle when plotted
against one another, as can be seen in \textbf{ (b) }. Both $I$ and $Q$ are digitalized to 8 bits of depth,
and are measured over several cycles, resulting in a thicker unit
circle. A phase random walk is obtiained by following
the method of equations \ref{eq:unwrap} and \ref{eq:diffandint}.
\textbf{(d)} shows the histogram of the different phase kicks that
are recovered from the phase random walk in (c).}
\end{figure}

With such values of photon numbers it is possible to measure clean
sinusoidal signals for the quadratures $I$ and $Q$ when they
are sampled at sample rates on the order of $\text{GHz}$, as illustrated
in Figure \ref{fig:SpaceWalksKicks}(a). This is in contrast to experiments
in which the vacuum field is one of the signals considered,
in which case the shot noise is the determinant origin of randomness \cite{Avesani,Zhou,Gabriel}.

\begin{figure}
\centering{}\includegraphics[width=1\columnwidth]{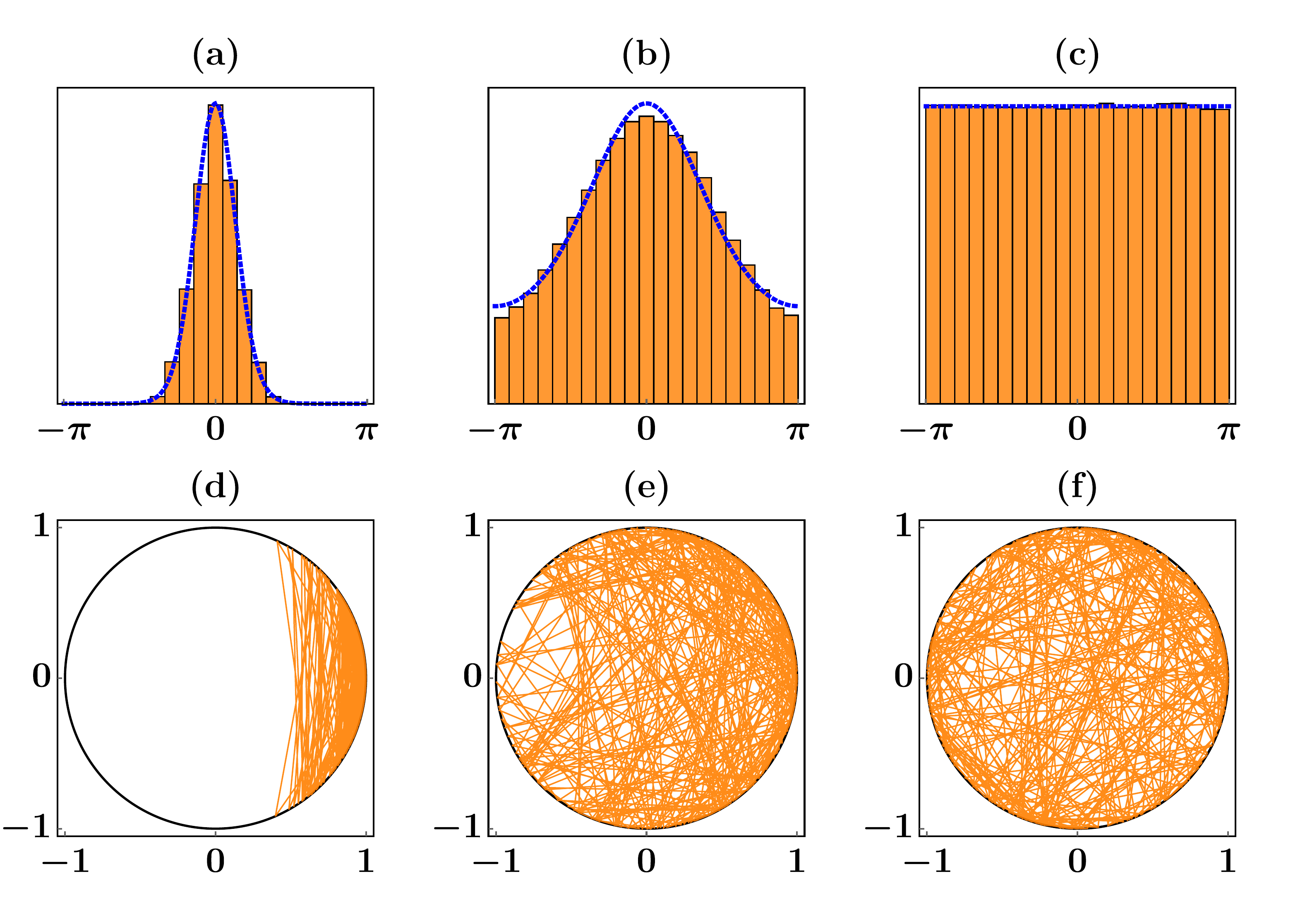}\caption{Histograms (with corresponding fits) and phase space trajectories
for three different sample rates, originating from 5 million point
sample measurements on a balanced coherent detector sampled at \textbf{(a,d)}
7.81MSamples/s, \textbf{(b,e)} 500kSamples/s, and \textbf{(c,f)} 156.25kSamples/s.
Only the first 300 points of each trajectory in the unit circle are
shown. The loss of correlation between successive kicks starts to
become noticeable once the sampling frequency is smaller than the
laser linewidths, i.e., at about 200kHz. All histograms are fit with
a von Mises distribution (equation \ref{eq:vonMises}) with variance
(a) $t/\overline{\tau_{c}}=0.172$, (b) $t/\overline{\tau_{c}}=1.78$,
and (c) $t/\overline{\tau_{c}}=7.50\times10^{5}$. (c) was additionally
fit with a uniform distribution, with a Goodness of fit Kolmogorov-Smirnov
test p-value of 0.56. \label{fig:histowool1}}
\end{figure}

With measurements in two complementary quadratures, i.e. by measuring
$I\equiv\left\langle x\left(0\right)\right\rangle $ and $Q\equiv\left\langle x\left(\pi/2\right)\right\rangle $,
it is possible have access to the complete phase dynamics of $\xi\left(t\right)$
directly. $I$ and $Q$ follow probability distributions which are
correlated and complementary. Having an additional quadrature provides
additional information on the value of the phase in the unit circle
instead of just one projection onto the axis, which has been the customary
technique for developing phase noise QRNGs which utilize randomness
extraction\cite{Mitchell,Abellan,Sun2017}. With this additional
information, it is possible to retreive a distribution which can be
made arbitrarily similar to a Uniform distribution stemming from quantum
randomness. With the information provided by the signals $I$ and
$Q$ and the unwrapping of the phase, we can reconstruct the phase
behavior of the laser interference: 
\begin{equation}
\Theta\left(t\right):=\text{unwrap}\left(\text{arg}\left(I+iQ\right)\right)=\Delta_{\omega}t+\xi\left(t\right).\label{eq:unwrap}
\end{equation}
It is possible to eliminate the main trend, $\Delta_{\omega}t$, to
obtain a phase variation which describes a random walk whose origin
are the spontaneous emission kicks inside of the laser cavities. This
is done by differentiating, subtracting the offset, and integrating
again, i.e.:

\begin{equation}
\xi\left(t\right)=\int\left(\frac{d\Theta}{dt}-\Delta_{\omega}\right)dt.\label{eq:diffandint}
\end{equation}
The random values are obtained by further differentiating this phase
($d\xi/dt$), wrapping it to the range $\left]-\pi,\pi\right]$ and
discretizing the possible outcomes to 8 bits. With this processing
technique it is possible to recover the kicks in the phase distribution,
following equation \ref{eq:vonMises}. The measurement of such phase
distributions is shown in Figure \ref{fig:histowool1}. With this
technique in the measured experimental data, we are also able to recover
directly the particular and different random trajectories of the phase.

As predicted by Equation \ref{eq:vonMises} and shown in Figure \ref{fig:histowool1},
phases start becoming uniformly distributed as time progresses. This
can be seen on the histograms, which widen and flatten as the time
between successive measurements becomes larger and thus converge to
uniform distributions of uncorrelated data sets. The autocorrelation
coefficient (as defined in \cite{ParkBook}) of different data points
never exceeds $10^{-2}$, and a figure of the autocorrelation for up to $2.5\times10^{6}$ samples is shown in the Supplementary information.

For random number generation, the measurements of $I$ and $Q$ are performed at 156.25 kSamples/s
and digitalized at an 8-bit depth. Values of $\theta$ are saved as
8-bit values to reach a random number generator speed of 1.25MBits/s. We have tested the generation of uniform random numbers in a binary
file of 21.21 Gbytes of data $(\approx2^{37.3}\text{ bits}$), corresponding
to $2.12\times10^{10}$ measurements (as every measurement provides
one byte) gathered in the experimental setup shown in Figure \ref{fig:EquivalenceFigure}
d). We have used three statistical suites generated for the verification
of Random Number Generators: Robert G. Brown's dieharder \cite{Dieharder},
United States' National Institute of Standards and Technology (NIST)'s
Statistical Testing Suite \cite{NIST}, and Pierre L'Ecuyer's TestU01's
Alphabit testing battery, specifically designed for the testing of
hardware Random Number Generators \cite{TestU01}. The random data
sequence passes 304 out of the 311 random tests applied to it. The
results of the dieharder tests are presented in the Supplementary Information C. A possible application for such a generator is the
provision of true, quantum random data to feed the extraction of faster
generators. Faster sources could be achieved by larger laser bandwidths,
but this is a problem to be addressed in future work.

\section{Conclusions}

We have demonstrated a quantum random number generator based on the
extraction of the random phase difference between two laser beams.
At slow speeds with high photon numbers, the phase noise is the dominant
contribution to the random character of the interference of two lasers
signals. The value of the phase is obtained by measuring two quadratures
of the electric field of the two interfering lasers with a coherent
detector. This randomness can be exploited to produce random numbers
which do not require randomness extraction to pass almost all of the
standard randomness tests designed for Random Number Generators, thus
providing a reliable source of true randomness that can be used for
further calibration of faster randomness extractors.

The simplicity of the processing techniques used and the availability
and sturdiness of the components in telecom laboratories makes it
easy to produce trusted random numbers for cryptographic purposes.
The technique presented can be further improved by measuring the randomness
of the phase with broader sources, such as Erbium Doped Fiber Amplifiers
(EDFAs).

The usage of narrowband sources such as the ones considered in this
paper are a drawback to the speed of the key generation, as the coherence
times of these lasers are considerably slow. However, it is important
to remark that even though certain applications might require high-speed
random number generation, in excess of a few Gbit/s, this is not a
restrictive requisite in all possible applications. In important cases,
the use of mature and easily accessible technology, the robustness
or even the availability of the components needed for random number
generation can be far more important considerations to take into account
than the RNG speed\cite{Chen}. 
\begin{acknowledgments}
JRA acknowledges V. Vicu\~{n}a, M. Mitchell, M. Avesani, S. Zubieta,
C. Leung and E. Kassa for useful discussions, as well as Fundaci\'{o}
Catalunya-La Pedrera's ICFO Summer Fellows and the Europhotonics POESII
Master. SS acknowledges MINECO FPI-BES-2015-074302. JAL acknowledges
ALLIANCE (TEC2017-90034-C2-2-R), project co-funded by FEDER. JPT acknowledges
financial support from Fundaci\'{o} Cellex, from the Government of
Spain through the Severo Ochoa Programme for Centres of Excellence
in R\&D (SEV- 2015-0522), and from Generalitat de Catalunya under
the programs ICREA Academia and CERCA. 
\end{acknowledgments}

\end{document}


\title{Quantum random number generation by coherent detection of phase noise\\
Supplementary information}
\maketitle

\subsection{Discretization of distribution}

\begin{figure}[H]
\begin{centering}
\includegraphics[width=0.5\columnwidth]{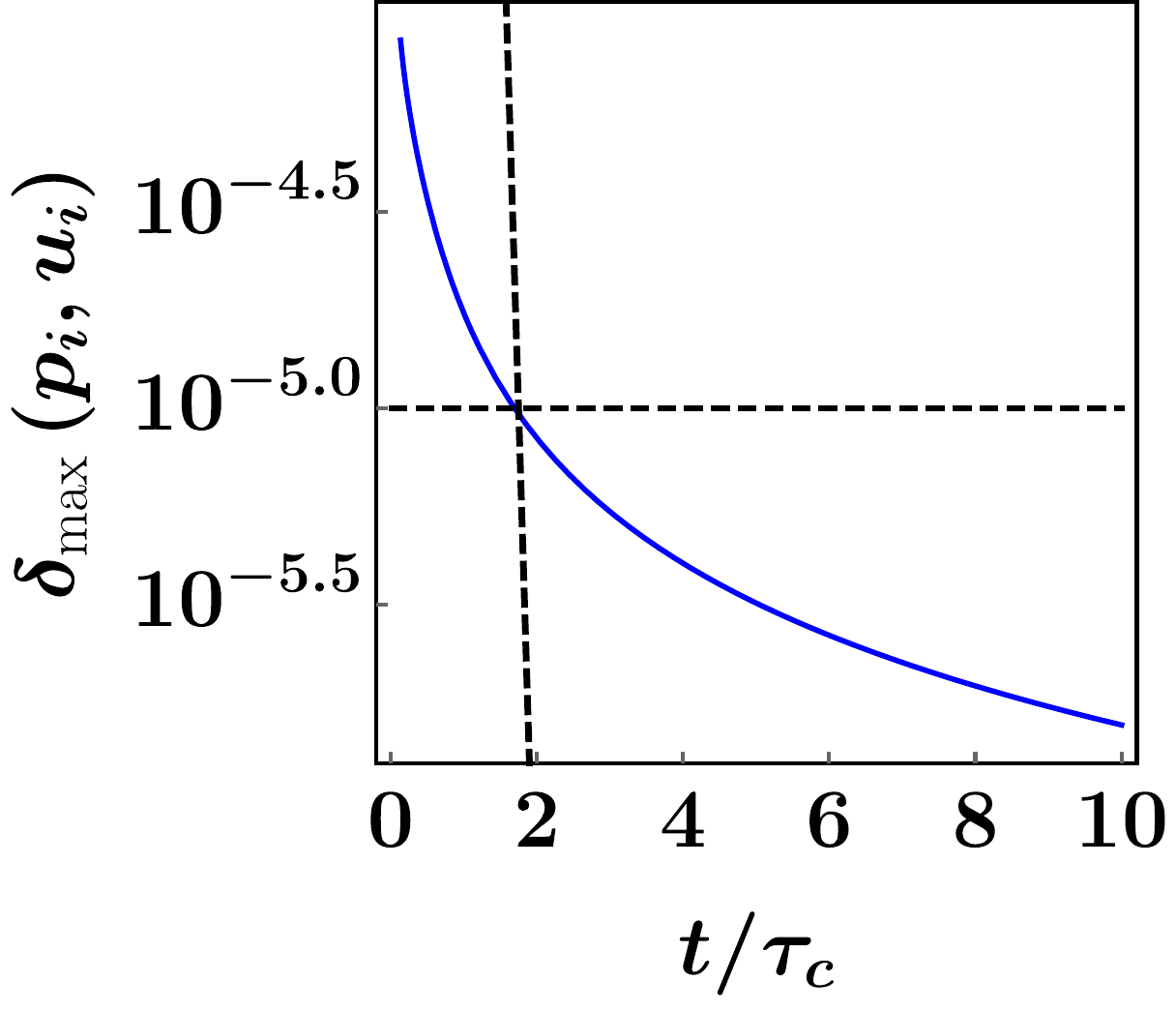} 
\par\end{centering}
\caption{Maximum difference between the discretized values of a von Mises distribution
of variance $t/\tau_{c}$ and a uniform distribution. Here, the discretization
is performed in $2^{8}$ values. \label{fig:DistanceUnif}}
\end{figure}

If the measurements of $\theta$ are discretized, the probability
distribution to which we have access will be 
\begin{equation}
p_{i}=\int_{-\pi+i\delta}^{-\pi+i\left(\delta+1\right)}P\left(\xi\left(t\right)=\theta\right)d\theta
\end{equation}
where $\delta=\frac{2\pi}{2^{k}}$, with $k$ being the bit depth
of the possible values in which $\theta$ can be discretized, and
$i\in\left\{ 0,2^{k}-1\right\} $. With $t/\overline{\tau_{c}}>2$
and $k=8$, the maximum difference between $p_{i}$ and $u_{i}$,
the value of a uniform distribution, does not exceed $10^{-5}$, i.e.,
it would take on average $10^{5}$ values to observe a difference
between the observed value and a uniform distribution, as shown in
Figure \ref{fig:DistanceUnif}.

\pagebreak{}

\subsection{Autocorrelation data}

\begin{figure}[H]
\begin{centering}
\includegraphics[width=1\columnwidth]{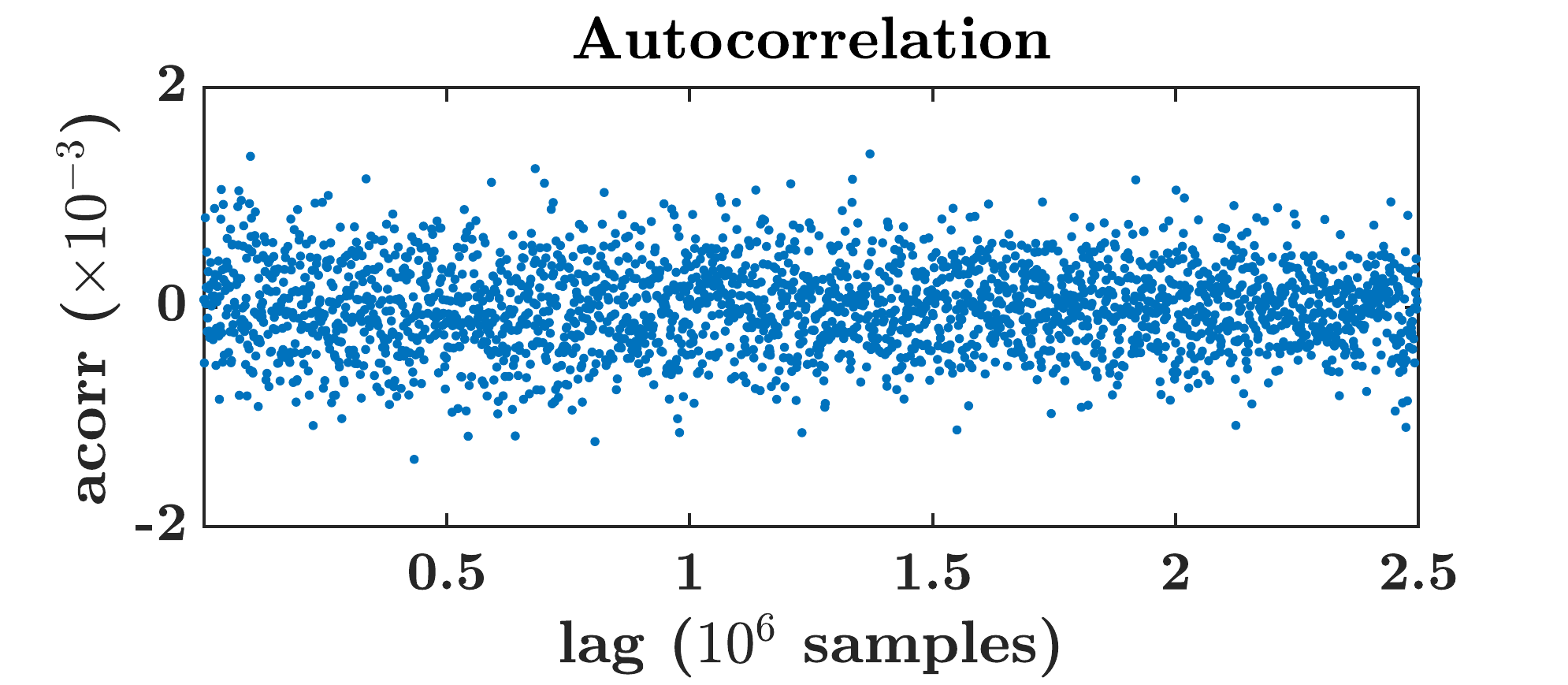} 
\par\end{centering}
\caption{Autocorrelation coefficients when the phase is being measured at rates
of 156.25 kSamples/s.\label{fig:autocorr-1}}
\end{figure}

Figure \ref{fig:autocorr-1} shows the autocorrelation coefficient
values with a lag $d$, 
\begin{equation}
K_{\Theta\Theta}\left(d\right)=\frac{\sum_{i=1}^{N-d}\left(\Theta_{i}-\text{mean}\left(\Theta\right)\right)\left(\Theta_{i+d}-\text{mean}\left(\Theta\right)\right)}{\text{var}(\Theta)},
\end{equation}
of a measured string of $N=5\times10^{6}$ random phases $\Theta$.
The autocorrelation for lag 0, which is by definition 1, is not shown.
The measurement of this autocorrelation coefficients is shown as measured
directly from the experiment, without any postprocesing apart from
the ones explained to retrieve the phase values. The autocorrelations
with $d>0$ are normally distributed with a mean zero and a standard
deviation of $3.87\times10^{-4}$.\\

\subsection{Statistical test results}

We have tested the generation of uniform random numbers in a binary
file of 21.21 Gbytes of data $(\approx2^{37.3}\text{ bits}$), corresponding
to $2.12\times10^{10}$ measurements. We have used three statistical
suites implemented for the verification of Random Number Generators:
Robert G. Brown's dieharder , passing 115 of the 117 evaluated tests,
United States' National Institute of Standards and Technology (NIST)'s
Statistical Testing Suite, passing 175 of the 177 evaluated tests,
and Pierre L'Ecuyer's TestU01's Alphabit testing battery, passing
14 of the 17 evaluated tests. The results have been grouped under
the smallest p-value.\\

\begin{table*}
\caption{\label{tab:table1}Top Left: TestU01's Alphabit test results. Top
Right: Dieharder test results. 21.21GB were assessed. Bottom: NIST
test results with proportion of passing sequences. 1000 sequences
of 20 million bits were assessed. The minimum pass rate for each test
is approximately of 980/1000 binary sequences, except for the random
excursion tests, which have a passing rate of approximately = 850/868
samples.\label{tab:testu01}}

\begin{tabular}{|llllll|llllll|}
\multicolumn{1}{l}{} &  &  &  &  & \multicolumn{1}{l}{} &  &  &  &  &  & \multicolumn{1}{l}{}\tabularnewline
\hline 
\textbf{Dieharder}  &  &  &  &  &  &  & \textbf{Alphabit}  &  &  &  & \tabularnewline
\hline 
 &  &  &  &  &  &  &  &  &  &  & \tabularnewline
\textbf{\textsc{Statistical test}}  &  & \textbf{\textsc{p-value}}  &  & \textbf{\textsc{assessment}}  &  &  & \textbf{\textsc{Statistical test}}  &  & \textbf{\textsc{p-value}}  &  & \textbf{\textsc{assessment}}\tabularnewline
 &  &  &  &  &  &  &  &  &  &  & \tabularnewline
diehard birthdays  &  & 0.31789570  &  & passed  &  &  & MultinomialBitsOver, L=2  &  & 0.20  &  & passed\tabularnewline
diehard operm5  &  & 0.78749679  &  & passed  &  &  & MultinomialBitsOver, L=4  &  & 0.11  &  & passed\tabularnewline
diehard rank 32x32  &  & 0.94675484  &  & passed  &  &  & MultinomialBitsOver, L=8  &  & 3.3e-31  &  & weak\tabularnewline
diehard rank 6x8  &  & 0.52261092  &  & passed  &  &  & MultinomialBitsOver, L=16  &  & 0.76  &  & passed\tabularnewline
diehard bitstream  &  & 0.73687190  &  & passed  &  &  & HammingIndep, L=16  &  & 0.17  &  & passed\tabularnewline
diehard opso  &  & 0.78549330  &  & passed  &  &  & HammingIndep, L=32  &  & 0.08  &  & passed\tabularnewline
diehard oqso  &  & 0.16121539  &  & passed  &  &  & HammingCorr, L=32  &  & 0.70  &  & passed\tabularnewline
diehard dna  &  & 0.32705856  &  & passed  &  &  & RandomWalk1 H, L=64  &  & 5.6e-4  &  & weak\tabularnewline
diehard count 1s str  &  & 0.33775770  &  & passed  &  &  & RandomWalk1 M, L=64  &  & 0.09  &  & passed\tabularnewline
diehard count 1s byt  &  & 0.47672743  &  & passed  &  &  & RandomWalk1 J, L=64  &  & 0.58  &  & passed\tabularnewline
diehard parking lot  &  & 0.95085394  &  & passed  &  &  & RandomWalk1 R, L=64  &  & 0.75  &  & passed\tabularnewline
diehard 2dsphere  &  & 0.40464477  &  & passed  &  &  & RandomWalk1 C, L=64  &  & 0.97  &  & passed\tabularnewline
diehard 3dsphere  &  & 0.83863434  &  & passed  &  &  & RandomWalk1 H, L=320  &  & 2.8e-3  &  & passed\tabularnewline
diehard squeeze  &  & 0.68552263  &  & passed  &  &  & RandomWalk1 M, L=320  &  & 5.0e-4  &  & weak\tabularnewline
diehard sums  &  & 0.16894560  &  & passed  &  &  & RandomWalk1 J, L=320  &  & 0.40  &  & passed\tabularnewline
diehard runs  &  & 0.77992792  &  & passed  &  &  & RandomWalk1 R, L=320  &  & 0.97  &  & passed\tabularnewline
diehard craps  &  & 0.12273374  &  & passed  &  &  & RandomWalk1 C, L=320  &  & 4.8e-3  &  & passed\tabularnewline
marsaglia tsang gcd  &  & 0.86153798  &  & passed  &  &  &  &  &  &  & \tabularnewline
sts monobit  &  & 0.63904175  &  & passed  &  &  &  &  &  &  & \tabularnewline
sts serial  &  & 0.99926271  &  & weak  &  &  &  &  &  &  & \tabularnewline
rgb minimum distance  &  & 0.00804042  &  & passed  &  &  &  &  &  &  & \tabularnewline
rgb permutations  &  & 0.37608574  &  & passed  &  &  &  &  &  &  & \tabularnewline
rgb lagged sum  &  & 0.00375789  &  & weak  &  &  &  &  &  &  & \tabularnewline
rgb kstest test  &  & 0.84371162  &  & passed  &  &  &  &  &  &  & \tabularnewline
dab bytedistrib  &  & 0.00630053  &  & passed  &  &  &  &  &  &  & \tabularnewline
dab dct  &  & 0.13385443  &  & passed  &  &  &  &  &  &  & \tabularnewline
dab filltree  &  & 0.40296775  &  & passed  &  &  &  &  &  &  & \tabularnewline
dab filltree2  &  & 0.26179356  &  & passed  &  &  &  &  &  &  & \tabularnewline
dab monobit2  &  & 0.90885239  &  & passed  &  &  &  &  &  &  & \tabularnewline
\hline 
\end{tabular}\\
 \bigskip{}
 \centering{}%
\begin{tabular}{lllllllllllllllll|lllllll|lllllllllllll}
\cline{18-24} \cline{19-24} \cline{20-24} \cline{21-24} \cline{22-24} \cline{23-24} \cline{24-24} 
 &  &  &  &  &  &  &  &  &  &  &  &  &  &  &  &  & \textbf{NIST}  &  &  &  &  &  &  &  &  &  &  &  &  &  &  &  &  &  &  & \tabularnewline
\cline{18-24} \cline{19-24} \cline{20-24} \cline{21-24} \cline{22-24} \cline{23-24} \cline{24-24} 
 &  &  &  &  &  &  &  &  &  &  &  &  &  &  &  &  &  &  &  &  &  &  &  &  &  &  &  &  &  &  &  &  &  &  &  & \tabularnewline
 &  &  &  &  &  &  &  &  &  &  &  &  &  &  &  &  & \textbf{\textsc{Statistical test}}  &  & \textbf{\textsc{p-value}}  &  & \textbf{\textsc{prop}}  &  & \textbf{\textsc{assessment}}  &  &  &  &  &  &  &  &  &  &  &  &  & \tabularnewline
 &  &  &  &  &  &  &  &  &  &  &  &  &  &  &  &  &  &  &  &  &  &  &  &  &  &  &  &  &  &  &  &  &  &  &  & \tabularnewline
 &  &  &  &  &  &  &  &  &  &  &  &  &  &  &  &  & Frequency  &  & 0.439122  &  & 989/1000  &  & passed  &  &  &  &  &  &  &  &  &  &  &  &  & \tabularnewline
 &  &  &  &  &  &  &  &  &  &  &  &  &  &  &  &  & BlockFrequency  &  & 0.753844  &  & 983/1000  &  & passed  &  &  &  &  &  &  &  &  &  &  &  &  & \tabularnewline
 &  &  &  &  &  &  &  &  &  &  &  &  &  &  &  &  & CumulativeSums  &  & 0.311542  &  & 991/1000  &  & passed  &  &  &  &  &  &  &  &  &  &  &  &  & \tabularnewline
 &  &  &  &  &  &  &  &  &  &  &  &  &  &  &  &  & CumulativeSums  &  & 0.903338  &  & 993/1000  &  & passed  &  &  &  &  &  &  &  &  &  &  &  &  & \tabularnewline
 &  &  &  &  &  &  &  &  &  &  &  &  &  &  &  &  & Runs  &  & 0.424453  &  & 993/1000  &  & passed  &  &  &  &  &  &  &  &  &  &  &  &  & \tabularnewline
 &  &  &  &  &  &  &  &  &  &  &  &  &  &  &  &  & LongestRun  &  & 0.250558  &  & 983/1000  &  & passed  &  &  &  &  &  &  &  &  &  &  &  &  & \tabularnewline
 &  &  &  &  &  &  &  &  &  &  &  &  &  &  &  &  & Rank  &  & 0.701366  &  & 992/1000  &  & passed  &  &  &  &  &  &  &  &  &  &  &  &  & \tabularnewline
 &  &  &  &  &  &  &  &  &  &  &  &  &  &  &  &  & FFT  &  & 0.450297  &  & 988/1000  &  & passed  &  &  &  &  &  &  &  &  &  &  &  &  & \tabularnewline
 &  &  &  &  &  &  &  &  &  &  &  &  &  &  &  &  & NonOverlappingTemplate  &  & 0.526105  &  & 979/1000  &  & weak  &  &  &  &  &  &  &  &  &  &  &  &  & \tabularnewline
 &  &  &  &  &  &  &  &  &  &  &  &  &  &  &  &  & OverlappingTemplate  &  & 0.000000  &  & 971/1000  &  & failed  &  &  &  &  &  &  &  &  &  &  &  &  & \tabularnewline
 &  &  &  &  &  &  &  &  &  &  &  &  &  &  &  &  & Universal  &  & 0.514124  &  & 988/1000  &  & passed  &  &  &  &  &  &  &  &  &  &  &  &  & \tabularnewline
 &  &  &  &  &  &  &  &  &  &  &  &  &  &  &  &  & ApproximateEntropy  &  & 0.581082  &  & 991/1000  &  & passed  &  &  &  &  &  &  &  &  &  &  &  &  & \tabularnewline
 &  &  &  &  &  &  &  &  &  &  &  &  &  &  &  &  & RandomExcursions  &  & 0.381439  &  & 860/868  &  & passed  &  &  &  &  &  &  &  &  &  &  &  &  & \tabularnewline
 &  &  &  &  &  &  &  &  &  &  &  &  &  &  &  &  & RandomExcursionsVariant  &  & 0.252554  &  & 857/868  &  & passed  &  &  &  &  &  &  &  &  &  &  &  &  & \tabularnewline
 &  &  &  &  &  &  &  &  &  &  &  &  &  &  &  &  & Serial  &  & 0.534146  &  & 990/1000  &  & passed  &  &  &  &  &  &  &  &  &  &  &  &  & \tabularnewline
 &  &  &  &  &  &  &  &  &  &  &  &  &  &  &  &  & LinearComplexity  &  & 0.955835  &  & 992/1000  &  & passed  &  &  &  &  &  &  &  &  &  &  &  &  & \tabularnewline
\cline{18-24} \cline{19-24} \cline{20-24} \cline{21-24} \cline{22-24} \cline{23-24} \cline{24-24} 
 &  &  &  &  &  &  &  &  &  &  &  &  &  &  &  & \multicolumn{1}{l}{} &  &  &  &  &  &  & \multicolumn{1}{l}{} &  &  &  &  &  &  &  &  &  &  &  &  & \tabularnewline
 &  &  &  &  &  &  &  &  &  &  &  &  &  &  &  & \multicolumn{1}{l}{} &  &  &  &  &  &  & \multicolumn{1}{l}{} &  &  &  &  &  &  &  &  &  &  &  &  & \tabularnewline
\end{tabular}
\end{table*}